\documentclass[
      ,final
   ]
   {aipproc}

\layoutstyle{6x9}

\usepackage{epsf}

\def\eqn#1{eq.~(\ref{#1})}
\def\Eqn#1{Equation~(\ref{#1})}

\def\sect#1{section~{\ref{#1}}}

\def\fig#1{fig.~{\ref{#1}}}
\def\Fig#1{Figure~{\ref{#1}}}

\def\tree{{\rm tree}}

\def\tlambda{{\tilde\lambda}}
\def\e{\epsilon}
\def\ve{\varepsilon}
\def\sandpp#1.#2.#3{%
\left\langle\smash{#1}{\vphantom1}^{+}\right|{#2}%
\left|\smash{#3}{\vphantom1}^{+}\right\rangle}
\def\sandpm#1.#2.#3{%
\left\langle\smash{#1}{\vphantom1}^{+}\right|{#2}%
\left|\smash{#3}{\vphantom1}^{-}\right\rangle}
\def\sandmp#1.#2.#3{%
\left\langle\smash{#1}{\vphantom1}^{-}\right|{#2}%
\left|\smash{#3}{\vphantom1}^{+}\right\rangle}
\def\sandmm#1.#2.#3{%
\left\langle\smash{#1}{\vphantom1}^{-}\right|{#2}%
\left|\smash{#3}{\vphantom1}^{-}\right\rangle}
\def\spab#1.#2.#3{\sandmm#1.#2.#3}
\def\spba#1.#2.#3{\sandpp#1.#2.#3}
\def\spaa#1.#2.#3.#4{\sandmp#1.{#2#3}.#4}
\def\spbb#1.#2.#3.#4{\sandpm#1.{#2#3}.#4}
\def\spa#1.#2{\left\langle#1\,#2\right\rangle}
\def\spb#1.#2{\left[#1\,#2\right]}
\def\spash#1.#2{\vphantom{\hat K}\spa{\smash{#1}}.{\smash{#2}}}
\def\spbsh#1.#2{\vphantom{\hat K}\spb{\smash{#1}}.{\smash{#2}}}
\newbox\SlashedBox
\def\slashed#1{\setbox\SlashedBox=\hbox{#1}
\hbox to 0pt{\hbox to 1\wd\SlashedBox{\hfil/\hfil}\hss}#1}
\def\hboxtosizeof#1#2{\setbox\SlashedBox=\hbox{#1}
\hbox to 1\wd\SlashedBox{#2}}

\newbox\charbox
\newbox\slabox
\def\s#1{{      
        \setbox\charbox=\hbox{$#1$}
        \setbox\slabox=\hbox{$/$}
        \dimen\charbox=\ht\slabox
        \advance\dimen\charbox by -\dp\slabox
        \advance\dimen\charbox by -\ht\charbox
        \advance\dimen\charbox by \dp\charbox
        \divide\dimen\charbox by 2
        \raise-\dimen\charbox\hbox to \wd\charbox{\hss/\hss}
        \llap{$#1$}
}}
\def\ksl{\s{k}}
\def\Ksl{\s{K}}
\def\lr{\leftrightarrow}
\def\Tr{{\rm Tr}}
\def\del{\partial}
\def\Neqfour{{\cal N} = 4}

\begin{document}

\noindent SLAC--PUB--11583
\hfill hep-ph/0512111

\title{Twistor String Theory and QCD%
\footnote{Talk presented at the International Europhysics Conference 
on High Energy Physics, Lisbon, Portugal, July, 2005.
Research supported by the US Department of Energy under contract
DE--AC02--76SF00515.}
}



\author{Lance J. Dixon
}{address={Stanford Linear Accelerator Center,
              Stanford University,
             Stanford, CA 94309, USA}}

\begin{abstract}
I review recent progress in using twistor-inspired methods
to compute perturbative scattering amplitudes in gauge theory,
for application to collider physics. 
\end{abstract}

\maketitle

%
%


\section{Introduction}
\label{IntroSection}

In two years, a new window will open into physics at
the shortest distance scales.  The Large Hadron Collider
(LHC) will begin operation at CERN, providing proton-proton
collisions at 14~TeV center-of-mass energy, seven times
greater than the 2~TeV currently available 
in $p\bar{p}$ collisions at Fermilab's Tevatron.
The LHC luminosity should be a factor of 10
to 100 greater than the Tevatron's.  The combined rise 
in energy and luminosity will lead to a huge increase 
in the production of particles with masses in 
the range 100--1000~GeV, including electroweak vector bosons, 
top quarks, Higgs bosons, and of course new particles, 
representing physics beyond the Standard Model.

There are a lot of ideas for physics beyond the Standard Model,
many associated with the puzzle of electroweak symmetry breaking,
and with resolutions of the hierarchy problem --- why the weak scale
is so much smaller than the Planck scale.
Supersymmetry, for example, predicts a host of new particles
in the 100--1000~GeV mass range, including (in most versions) a 
stable dark matter candidate.  However, many other theories
--- new dimensions of space-time, new forces, {\it etc.} ---
often make qualitatively similar predictions.  How can we sort
out the predictions of these theories from each other, and
from the omnipresent Standard Model background at a hadron collider?

The short answer is that a thorough, quantitative understanding of 
both the new physics signals and the Standard Model backgrounds 
is required.  Much work has gone into these problems, stretching 
back over many decades.  This talk will focus on some recent developments, 
novel methods to help compute the backgrounds in particular, that have emerged 
since the Fall of 2003, when Witten introduced twistor string 
theory and explained its relevance
to perturbative QCD~\cite{WittenTopologicalString}.
In truth, the new methods have not yet had a direct phenomenological
impact, in terms of producing more accurate cross sections that have 
not previously been obtained in any other way.  But they have a 
lot of promise, and it should not be long before they do so.

\begin{figure}[t]
\centering
\includegraphics[height=.14\textheight]{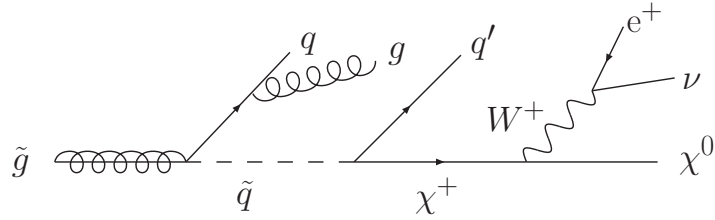}
\caption{Typical cascade decay of a gluino to two quarks,
a gluon, a $W$ boson which decays to two leptons,
and a neutralino.  Aside from the neutralino, all the final-state
particles are essentially massless.}
\label{CascadeFigure}
\end{figure}

What are some generic properties of the new physics signals?
Except for stable, neutral dark matter candidates,
the new massive particles typically decay into ``old'' Standard Model
particles: quarks, gluons, charged leptons and neutrinos,
photons, $W$s and $Z$s.  For example, in supersymmetry 
the superpartner of the gluon, the gluino, may be among
the heavier superparticles, yet still be copiously produced
at the LHC, due to its large adjoint color charge.
\Fig{CascadeFigure} shows a typical decay cascade, initiated by 
one of the two gluinos ($\tilde g$) in a pair-production event.
The quarks and gluons emerge as jets of hadrons.
The lightest superpartner, a neutralino ($\chi^0$), is stable and escapes
the detector.  The kinematic signatures of such events are not
always clean:  There can be a large number of observed particles 
(charged leptons or jets), and no invariant-mass bumps, because of the 
escaping neutralinos and possibly neutrinos (although there can be
kinematic edges).   The escaping neutralinos provide a missing 
transverse energy signal, but Standard Model production of 
$Z$ bosons, followed by $Z$ decays to neutrinos, can mimic this 
to some degree.

%

In order to maximize the potential for the discovery and interpretation
of new physics at the LHC,
we need to quantify the Standard Model backgrounds for processes 
that may contain several jets and (perhaps) a few electroweak bosons.
These processes are complex, so we should try to take into account
any simplifying features.  Notice that
the masses of the observed final-state particles in these reactions
({\it e.g.} in \fig{CascadeFigure}) are generally negligibly small 
in these reactions, except for the cases of the $W$, $Z$, or top quark.  
Even these particles immediately decay to essentially massless 
quarks or leptons, however.  So if we include the decay processes 
in the description of the event, every final-state particle is 
approximately massless.  We can also (usually) neglect the masses 
of the colliding partons (quarks and gluons).
In general, then, the backgrounds (and many signals)
require a detailed understanding of scattering amplitudes
for many ultra-relativistic (massless) particles -- especially 
the quarks and gluons of QCD.

Asymptotic freedom~\cite{GWP} allows us to compute such scattering
amplitudes as a perturbative expansion in the strong coupling constant
$\alpha_s(\mu)$, evaluated at a large momentum scale $\mu$ where it
is small.  For typical collider processes, $\mu$ could be of order
100--200 GeV, for which $\alpha_s(\mu) \approx 0.1$. 
One might expect that the leading-order terms in the expansion
(tree amplitudes) would suffice to get a 10\% uncertainty.
However, this is not the case for hadron collider cross sections; 
typical corrections from the next-to-leading order (NLO) terms in 
the $\alpha_s$ expansion are 30\% to 100\%.   There are several
possible reasons for the large corrections, depending on the process:   
there may be different scales involved,
leading to large logarithms of the ratio(s) of scales; 
new partonic subprocesses may first arise at NLO; 
the lowest-order process may have several factors 
of $\alpha_s$ in it; and so on.  In any event, a quantitative
description of collider events requires evaluation of cross
sections at NLO in QCD, which in turn requires, as input, one-loop
amplitudes as well as tree amplitudes.  If a precise 
evaluation is needed (below 10\% uncertainty), then the 
next-to-next-to-leading order terms, involving two-loop amplitudes, 
may also be required.

In principle, Feynman rules~\cite{FeynPT} are all we
need to evaluate the tree and loop amplitudes.  
In practice, however, although Feynman rules
are very general, applying to any local quantum field theory,
by the same token they are not optimized for the problems
at hand.  More efficient methods are available, which
make use of the extra symmetries (some hidden) of QCD.

\section{Transforming to twistor space}
\label{TwistorSection}

An easy way to see that there should be more efficient methods
out there is to notice that many QCD amplitudes are much 
simpler than expected.  For example, the tree-level amplitudes
for the scattering of $n$ gluons turn out to all vanish, 
if the helicities of the gluons (considered as outgoing particles) 
are either a) all the same, or b) all the same, except for one of 
opposite helicity.  Using parity, we can take the bulk of the gluons 
to have positive helicity, and write this vanishing relation as
\begin{equation}
A_n^\tree(1^\pm,2^+,3^+,\ldots,n^+) = 0 \,.
\label{susyvanish}
\end{equation}
This vanishing is somewhat mysterious from the point of view of Feynman
diagrams. On the other hand, 
it can be demonstrated simply using Ward identities 
arising from a secret supersymmetry that tree-level QCD amplitudes 
possess~\cite{SWI}.  This symmetry allows two of 
the gluons to be replaced by their superpartners, 
gluinos, which can be taken to be massless here.  Helicity 
conservation for the gluinos then implies the vanishing of
the amplitudes.

The first sequence of nonvanishing tree amplitudes has two gluons with
negative helicity, labelled by $j$ and $l$, say, and the rest of
positive helicity.  This sequence of maximally helicity-violating (MHV) 
amplitudes has an exceedingly simple form~\cite{ParkeTaylor,BGSixMPX},
\begin{equation}
A_n^{{\rm MHV}\,,jl}
\equiv
A_n^{\rm tree}(1^+,2^+,\ldots,j^-,\ldots,l^-,\ldots,n^+)
= i { {\spa{j}.{l}}^4 \over \spa1.2\spa2.3\cdots\spa{n}.1 } \,,
\label{PTAmps}
\end{equation}
in terms of spinor products $\spa{i}.{j}$ we shall define shortly.
\Eqn{PTAmps} is the expression for, not the full amplitude, but
rather a piece of it where the $n$ gluons have a definite cyclic 
ordering.  The full amplitude can be built out of 
permutations of such partial amplitudes, as reviewed for example
in ref.~\cite{TreeReview}.   Some of the structure of \eqn{PTAmps}
follows from supersymmetry, but not all.

To see much more of the structure, Witten~\cite{WittenTopologicalString}
transformed the amplitudes~(\ref{PTAmps}) from the traditional
momentum-space variables, into a twistor space invented 
by Penrose~\cite{Penrose}.  The twistor transform is a kind of
Fourier transform.  There are many examples where transforming
a problem into the right variables can expose its simplicity.
For example, if we measure the time dependence of the electric 
field $E(t)$ associated with the light emerging from some glowing sample of gas, 
we find a fairly unenlightening waveform.  However, if we use 
a spectrometer to measure instead the frequency (energy) 
spectrum of the light, that is, $E(\omega) = \int dt e^{i\omega t} E(t)$, 
we find spectral lines, which are clues toward decoding the 
structure of the emitting gas.  In in an analogous way, the twistor 
transform exposes certain lines on which QCD amplitudes
are localized or supported, thus revealing more of their structure,
and giving rise to new, more efficient ways to compute them.

Before describing the twistor transform, however,
we should discuss the spinor variables used 
in \eqn{PTAmps}, because they are well-suited for describing
scattering amplitudes for massless particles with spin,
and are the starting point for the twistor transform.  
Let $i=1,2,\ldots,n$ label the particles being scattered.
Usually, the four-momentum vectors $k_i^\mu$,
which transform under the spin-1 representation of the Lorentz group,
are used as the arguments of the amplitude, $A = A(k_i)$.
The relativistic invariants constructed out of these vectors
are the Lorentz inner products, or invariant masses, 
$s_{ij} = 2k_i\cdot k_j = (k_i+k_j)^2$,
which are equivalent in the massless case, $k_i^2 = 0$.
However, for massless particles with spin, it is better to 
``take the square root'' and use, instead of $k_i^\mu$, 
objects transforming as the spin-1/2 representation of the 
Lorentz group, namely the massless Dirac spinors associated 
with momentum $k_i$, $u_\pm(k_i)$, where the $\pm$ sign 
labels the helicity.  A shorthand notation for the two-component 
(Weyl) versions of these spinors is,
\begin{equation}
(\lambda_i)_\alpha \equiv \bigl[ u_+(k_i) \bigr]_\alpha,
\qquad
(\tlambda_i)_{\dot\alpha} \equiv \bigl[ u_-(k_i) \bigr]_{\dot\alpha}.
\end{equation}

We can always reconstruct the momenta from the spinors, using
the positive-energy projector for massless spinors, 
$u(k)\bar{u}(k) = \ksl$, or in two-component notation,
\begin{equation}
k_i^\mu (\sigma_\mu)_{\alpha\dot\alpha}
= (\ksl_i)_{\alpha\dot\alpha}
= (\lambda_i)_\alpha (\tlambda_i)_{\dot\alpha} \,.
\label{kfact}
\end{equation}
\Eqn{kfact} shows that a massless momentum vector, written
as a bi-spinor, is the product of a left-handed spinor
with a right-handed one.  

Instead of Lorentz inner products of momenta, 
$s_{ij} = 2k_i\cdot k_j$, we use spinor products, 
defined by
\begin{equation}
\spa{j}.{l}
= \ve^{\alpha\beta} (\lambda_j)_\alpha (\lambda_l)_\beta
= \bar{u}_-(k_j) u_+(k_l)\,,
\qquad
\spb{j}.{l}
= \ve^{\dot\alpha\dot\beta} (\tlambda_j)_{\dot\alpha} (\tlambda_l)_{\dot\beta}
= \bar{u}_+(k_j) u_-(k_l)\,,
\label{spinorproddef}
\end{equation}
where $\ve^{\alpha\beta}$ and $\ve^{\dot\alpha\dot\beta}$ are
antisymmetric tensors for $SU(2)$.  These products satisfy
\begin{equation}
\spa{i}.{j} \spb{j}.{i} 
= {1\over2} \Tr[ \ksl_i \ksl_j ] = 2k_i\cdot k_j = s_{ij} \,.
\label{spaspbeq}
\end{equation}
So they are just the square roots of the Lorentz inner products, up to a
phase $\phi$,
\begin{equation}
\spa{j}.{l}
= \sqrt{s_{jl}} e^{i\phi_{jl}} \,,
\qquad
\spb{j}.{l} = \pm \sqrt{s_{jl}} e^{-i\phi_{jl}} \,.
\label{spinorprodtwo}
\end{equation}

\begin{figure}[t]
\centering
\includegraphics[height=.13\textheight]{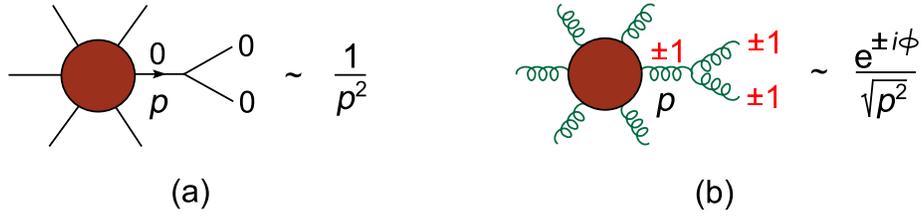}
\caption{Contrasting collinear behavior of amplitudes in 
(a) massless scalar $\phi^3$ theory, where there is no 
angular-momentum mismatch, and (b) massless gauge theory,
for example a gluon splitting into two gluons,
where there is always a mismatch.} 
\label{SGCollinearFigure}
\end{figure}

The utility of spinor variables for QCD amplitudes
was recognized already in the 
1980s~\cite{SpinorHelicity,BGSixMPX}.  They precisely capture
the ``square-root-plus-phase'' behavior of gauge theory amplitudes
as the momenta of two of the particles, $i$ and $j$, become collinear.
This behavior arises because the sum of the helicities of the final-state
particles is never equal to the helicity of the almost-on-shell 
intermediate particle, as illustrated in \fig{SGCollinearFigure}(b) 
for the case of a gluon splitting into two gluons, for which 
$\pm1\pm1 \neq \pm1$.
This mismatch in angular momentum along the collinear direction
lessens the singularity, from $1/s_{ij}$ (the behavior of
the scalar theory shown in \fig{SGCollinearFigure}(a))
to $1/\sqrt{s_{ij}}$.  
It also introduces a phase depending on the azimuthal
angle, which is conjugate to the angular momentum.
\Eqn{spinorprodtwo} shows that both characteristics
are captured by putting a spinor product in 
the denominator of the amplitude, explaining why the spinor products
are natural variables to use.  In other words, we should write
$A = A(\lambda_i,\tlambda_i)$ instead of $A = A(k_i)$.

%

Now we can describe the twistor 
transform~\cite{Penrose,WittenTopologicalString}.
It is a ``half'' Fourier transform, in which the right-handed spinors
$\lambda_i$ are left untouched, but each left-handed spinor
$\tlambda_i$ is exchanged for its Fourier conjugate variable $\mu_i$, 
defined by
\begin{equation}
\tlambda_{\dot\alpha} = i { \del \over \del \mu^{\dot\alpha} } \,,
\qquad
\mu^{\dot\alpha} = i { \del \over \del \tlambda_{\dot\alpha} } \,.
\label{mudef}
\end{equation}
(These relations are completely analogous to the standard
Fourier relation between momentum and position, 
$x = i \del/\del p$, $p = -i \del/\del x$.)
Since the spinors and their conjugates each have two components,
twistor space has four coordinates (for each external particle),
$(\lambda_1,\lambda_2,\mu^{\dot1},\mu^{\dot2})$.
However, because amplitudes are only defined up to a phase associated
with external states, two points in twistor space are equivalent 
if the four coordinates differ by a constant multiple $\xi$
(the complexification of the phase),
\begin{equation}
 (\lambda_1,\lambda_2,\mu^{\dot1},\mu^{\dot2})
 \equiv (\xi\lambda_1,\xi\lambda_2,\xi\mu^{\dot1},\xi\mu^{\dot2}).
\label{twcoord}
\end{equation}
So in fact (projective) twistor space is three-dimensional.

\begin{figure}
\centering
  \includegraphics[height=.20\textheight]{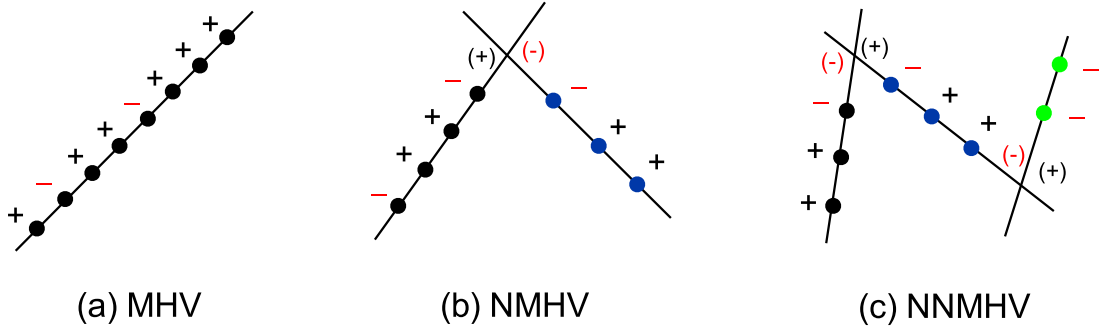}
  \caption{Tree amplitudes for $n$ gluons are supported on 
 networks of intersecting lines in twistor space.  The number of 
 lines is one fewer than the number of negative-helicity gluons.}
 \label{mhvtreefigure}
\end{figure}

What do amplitudes look like in this space?
We can compute them by Fourier transforming, just as
we would to take a wave-function from position-space to 
momentum-space~\cite{WittenTopologicalString},
\begin{equation}
  A(\lambda_i,\tlambda_i)\quad \Rightarrow \quad
  A(\lambda_i,\mu_i) 
\equiv \int \prod_{i=1}^n 
d\tlambda_i\, e^{i\mu_i \tlambda_i} 
  A(\lambda_i,\tlambda_i) \,.
\label{twtransform}
\end{equation}
The simplest cases to consider are the MHV amplitudes~(\ref{PTAmps}),
which contain only angle brackets ($\spa{i}.{j}$), and so
depend almost exclusively on the right-handed spinors $\lambda_i$, 
$A_n^{\rm MHV}(\lambda_i,\tlambda_i) \equiv A_n^{\rm MHV}(\lambda_i)$.
Their only dependence on the left-handed spinors
is through the usual momentum-conserving $\delta$-function
(which was implicit in \eqn{PTAmps}).
This factor can be written, using the identity
\begin{equation}
\delta^4(k) = \int d^4x \, \exp[ik\cdot x]
\label{deltaident}
\end{equation}
and \eqn{kfact}, as
\begin{equation}
\delta^4\Bigl( \sum_{i=1}^n k_i \Bigr)
= \int d^4x \, 
\exp\Bigl[ i x^{\alpha\dot\alpha}
\sum_{i=1}^n (\lambda_i)_\alpha (\tlambda_i)_{\dot\alpha} \Bigr] \,.
\label{deltafn}
\end{equation}
Then the transformed amplitudes are
\begin{eqnarray}
A_n^{{\rm MHV}}(\lambda_i,\mu_i)
 &=& \int \prod_{i=1}^n d\tlambda_i \exp[ i \mu_i \tlambda_i ]
  \int d^4x\ A_n^{{\rm MHV}}(\lambda_i)
 \exp[ i x \lambda_i \tlambda_i ] 
\nonumber\\
 &=& \int d^4x\ A_n^{{\rm MHV}}(\lambda_i)
     \int \prod_{i=1}^n d\tlambda_i
        \exp[ i (\mu_i + x \lambda_i ) \tlambda_i ]
\nonumber\\
&=& A_n^{{\rm MHV}}(\lambda_i) 
     \int d^4x \prod_{i=1}^n \delta( \mu_i + x \lambda_i ) \,.
\label{TwistorMHV}
\end{eqnarray}
The product of all the linear $\delta$-function constraints 
simply means that the amplitude is supported on a line 
in twistor space, as shown in \fig{mhvtreefigure}(a).

More complicated amplitudes can also be inspected.  The first
nonvanishing, non-MHV $n$-gluon amplitudes are the six-gluon
amplitudes with three positive and three negative helicities,
first computed, from 220 Feynman diagrams, in 1988~\cite{BGSixMPX}.
The simplest case, where the three positive helicities are adjacent,
is given by,
\begin{eqnarray}
A_6^\tree(1^+,2^+,3^+,4^-,5^-,6^-) &=& i \biggl[
{ ( \spb1.2 \spa4.5 \spab6.{(1+2)}.3 )^2 
\over s_{61} s_{12} s_{34} s_{45} s_{612} }
\nonumber\\
&&\hskip0cm
+ { ( \spb2.3 \spa5.6 \spab4.{(2+3)}.1 )^2 
\over s_{23} s_{34} s_{56} s_{61} s_{561} }
\nonumber\\
&&\hskip-4.5cm
+ { s_{123} \spb1.2 \spb2.3 \spa4.5 \spa5.6 
     \spab6.{(1+2)}.3 \spab4.{(2+3)}.1
  \over s_{12}  s_{23} s_{34} s_{45} s_{56} s_{61} } \biggr] \,.
\label{ApppmmmOLD}
\end{eqnarray}
where $s_{abc} \equiv (k_a+k_b+k_c)^2$ and 
$\spab{a}.{(b+c)}.{d} \equiv \bar{u}_-(k_a) (\ksl_b+\ksl_c) u_-(k_d)$.

The seven-gluon amplitudes were also computed around this 
time~\cite{BGKSeven}, using off-shell recursive 
methods~\cite{BGRecursive} to avoid dealing directly with
the 2,485 Feynman diagrams.  The explicit results in this case 
fill several pages.  Computing the twistor transform via
\eqn{twtransform} is rather difficult.  However, suppose one has
a guess for how the amplitudes are supported in twistor space,
for example that they are localized on some curve described 
by a polynomial equation $C(Z_i) = 0$, where 
$Z = (\lambda_1,\lambda_2,\mu^{\dot1},\mu^{\dot2})$.
Then it is relatively easy to check such a guess back
in spinor-space, where $C(Z_i)$ becomes a differential operator,
since $\mu_i = i \del/\del\tlambda_i$.  Applying $C(Z_i)$
to $A(\lambda_i,\tlambda_i)$, if the result vanishes identically
then the amplitude is supported on the curve; that is, either $C=0$
or else $A=0$.

This method was used last year by Cachazo, Svr\v{c}ek and
Witten~\cite{WittenTopologicalString,CSWI,CSWII} 
to build up evidence for the picture illustrated in \fig{mhvtreefigure}.
Scattering amplitudes for $n$ gluons, of which $n_-$
have negative helicity, are localized in twistor space
on networks of intersecting lines, where the number of 
intersecting lines is $n_- - 1$.  The MHV case, $n_-=2$,
was discussed above.  The next-to-MHV (NMHV) amplitudes with $n_-=3$, 
for example the six-gluon example in \eqn{ApppmmmOLD}, 
are sums of terms, each of which is supported on a pair 
of intersecting lines, as shown in \fig{mhvtreefigure}(b).
The partitioning of points among the lines can vary from term
to term.  Three intersecting lines are needed to describe the
next-to-next-to-MHV (NNMHV) amplitudes with $n_-=4$
(\fig{mhvtreefigure}(c)), and so on.

\section{MHV rules}
\label{MHVRulesSection}

\begin{figure}
\centering
  \includegraphics[height=.13\textheight]{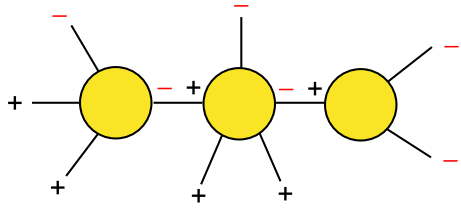}
  \caption{Example of an MHV diagram, corresponding to
  fig. 3(c).}
 \label{mhvdiagram}
\end{figure}

While the twistor structure shown in \fig{mhvtreefigure}
is extremely appealing, it does not directly
yield the numerical values of the amplitudes.  However,
Cachazo, Svr\v{c}ek and Witten~\cite{CSWI} also wrote down a 
set of diagrammatic ``MHV'' rules, which can be used in place of
Feynman rules to compute the amplitudes, and which make 
the twistor structure in \fig{mhvtreefigure} manifest.  Each MHV diagram
generates a term in the amplitude which has one of the possible
twistor structures, taking into account the possible partitionings
of points among the $(n_{-}-1)$ lines.
For example, the MHV diagram in \fig{mhvdiagram}, for an amplitude
with $n_-=4$, corresponds to the twistor structure in 
\fig{mhvtreefigure}(c).  The helicities of internal, as well
as external, gluons are labeled by $\pm$ in the diagram.  
Each vertex must have exactly two negative-helicity gluons attached to it,
but it can have an arbitrary number of positive-helicity gluons,
just like the MHV amplitude~(\ref{PTAmps}).   In fact,
the rule for this MHV vertex (the complex number associated with it)
is given by \eqn{PTAmps}, with a simple prescription for continuing
intermediate legs off shell.  The rule for an internal line is 
a factor of $i/p^2$, much like a scalar propagator.
For processes with a large number of gluons, there are considerably
fewer MHV diagrams than Feynman diagrams, because many Feynman subdiagrams
get lumped into single MHV vertices.  Also, the algebra
required to evaluate each diagram is considerably simpler than for the 
typical Feynman diagram, because there is no tangle of Lorentz indices
to follow.

Because the MHV rules are so efficient, they were quickly
generalized to a more general set of processes of interest
in the context of LHC signals and backgrounds:
tree-level QCD amplitudes containing massless external
fermions as well as gluons~\cite{ExtFerm};
those with a Higgs boson, which couples to gluons via 
$H\Tr(G_{\mu\nu}G^{\mu\nu})$ in the large $m_t$ limit~\cite{ExtHiggs}; 
and amplitudes including one or more electroweak vector bosons 
in addition to massless quarks and gluons~\cite{ExtVector}.  
A set of scalar-type rules for QCD with massive quarks ({\it e.g.}
the top quark) was also produced, starting directly from the QCD
Lagrangian~\cite{SchwinnWeinzierl}.

\section{Twistor structure at one loop}
\label{TwistorLoopSection}

In parallel with the extension of tree-level MHV rules to different
processes, the twistor structure of one-loop amplitudes
began to be investigated~\cite{CSWII}.
For multi-particle processes, one-loop amplitudes are much more
intricate than tree amplitudes.  Their twistor structure 
is also complicated by a ``holomorphic anomaly''~\cite{HoloAnomaly}, 
in which derivatives from $\mu_i = i \del/\del\tlambda_i$
act near singular regions of the loop integration.
For these reasons, it has proven simpler to proceed by first representing
amplitudes as linear combinations of various types of basic one-loop
integrals --- boxes, triangles, bubbles, {\it etc.} ---
and then examining the twistor structure of the coefficients of these
integrals.  

The simplest situation to consider is a
``toy model'' for perturbative QCD, namely its maximally supersymmetric
cousin, $\Neqfour$ super-Yang-Mills theory.  In this theory,
the coefficients of the triangle and bubble integrals all 
vanish, reducing the problem to that of determining the
coefficients of box integrals~\cite{Neq4MHV}.
These coefficients can be found quite
readily~\cite{HoloAnomaly,NMHVSeven,BCFGenUnitarity,Neq4NMHV}
by inspecting 
either standard two-particle unitarity
cuts~\cite{Cutting,Neq4MHV,LoopReview},
or (more efficiently) generalized cuts~\cite{ELOP,Zqggq} 
where four propagators are held
open~\cite{BCFGenUnitarity}.

\begin{figure}
\centering
  \includegraphics[height=.26\textheight]{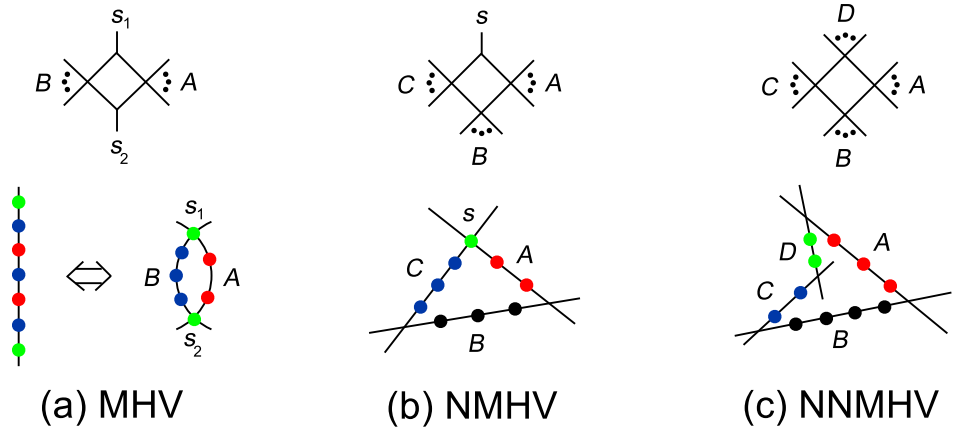}
\caption{Twistor structure of box integral coefficients for
one-loop amplitudes in $\Neqfour$ supersymmetric Yang-Mills theory.
For each series of amplitudes, (a) MHV ($n_-=2$), (b) NMHV ($n_-=3$), 
and (c) NNMHV ($n_-=4$), the type of box integral having the simplest 
nonvanishing coefficient is depicted at the top, and the localization
of those coefficients in twistor space is shown at the bottom.}
\label{looptwistfigure}
\end{figure}

The resulting twistor structure~\cite{NMHVSeven,Neq4NMHV,BCFCoplanarity}
is illustrated in \fig{looptwistfigure}.
In the MHV case shown in \fig{looptwistfigure}(a), the only
nonvanishing box coefficients are those where two of the external
momenta for the scattering amplitude, $s_1$ and $s_2$, 
are also momenta for the box integral;
the remaining external momenta are partitioned into two diagonally
opposite clusters, $A$ and $B$.  This integral is referred to
as a two-mass box, because the clustered momenta $K_A = \sum_{i\in A} k_i$
and $K_B = \sum_{i\in B} k_i$ are massive, $K^2_{A,B} \neq 0$.
The coefficient of the two-mass box~\cite{Neq4MHV} is just the 
MHV tree amplitude~(\ref{PTAmps}), 
which is localized on a single line in twistor space 
(see \fig{mhvtreefigure}(a)). In \fig{looptwistfigure}(a), 
the single line has been redrawn as a pair of lines intersecting 
in two points, $s_1$ and $s_2$, to make its appearance consistent 
with an ``MHV rules'' approach to one-loop amplitudes~\cite{BST},
and with the pattern found for more negative-helicity gluons. 
(Just as in Euclidean space, a pair of straight lines intersecting in 
two points in twistor space is the same as a single line.)
In the NMHV case shown in \fig{looptwistfigure}(b), the simplest
nonvanishing box coefficients are generically those of the three-mass box
integral, for which three of the legs, $A$, $B$, and $C$, 
represent clusters of momenta from the scattering amplitude, 
and only one, $s$, is an individual scattering momentum.  
These coefficients have a planar twistor structure, consisting
of three intersecting lines, and the leg $s$ sits at one
of the intersections~\cite{NMHVSeven,BCFCoplanarity,Neq4NMHV}.
For the NNMHV case in \fig{looptwistfigure}(c), the
four-mass box coefficients have the nonplanar ring structure
shown~\cite{Neq4NMHV}.  In general, as in the tree case, 
\fig{mhvtreefigure}, one-loop box coefficients are supported
on networks of lines, but the lines are joined into rings to
match the loop topology.
Similar structures have been found for coefficients of integrals 
in gauge theories with ${\cal N} < 4$
supersymmetries~\cite{NeqoneTwistor}.

\section{What is a twistor string?}
\label{TwistorStringSection}

I have been remiss in titling this talk ``Twistor String Theory
and QCD,'' without saying anything yet about what twistor
string theory is, or how it is related to the more phenomenological
developments just outlined.
In fact, I won't describe twistor string theory at any length, 
but I would like to briefly contrast it with ordinary string 
theory, from the perspective of methods for computing gauge theory amplitudes.

\begin{figure}
\centering
  \includegraphics[height=.21\textheight]{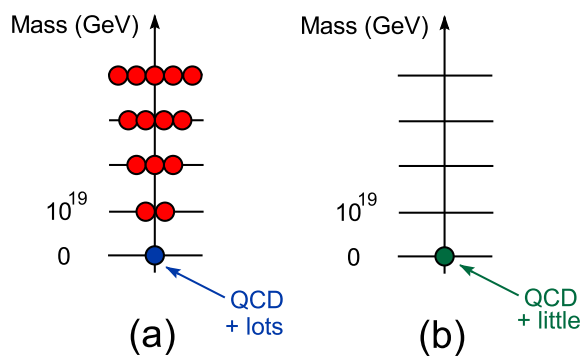}
  \caption{(a) Typical spectrum of particles in ordinary string theory.
  (b) Typical spectrum in (topological) twistor string theory.}
 \label{StringSpectrum}
\end{figure}

An ordinary string is an extended object which moves in
space-time.  Different physical vibrations of the string are associated
with different particle states.  The higher the harmonic, the more
massive the particle; indeed, there is an infinite tower of 
ultra-heavy particles, as well as a set of massless ones.
One of the massless particles is always the graviton.
Because the one energy scale in gravity is the Planck mass, 
$M_P \approx 10^{19}$~GeV, this sets the scale for the ultra-heavy
masses (unless certain extra dimensions happen to have a large size),
as shown in \fig{StringSpectrum}(a).
Also, the massless spectrum can be relatively complicated ---
several gauge groups, matter fields transforming in various ways,
and so on.

In the early 1990s, this type of string theory was adapted 
by Bern and Kosower into a tool to compute one-loop QCD 
amplitudes~\cite{BernKosower}.  It worked pretty well.  
The first computations of the helicity amplitudes for four-gluon
scattering in QCD used this technique~\cite{BernKosower}.
(The unpolarized cross sections were computed earlier via more
traditional methods~\cite{EllisSexton}.)
The five-gluon helicity amplitudes were also computed via
the string-based method~\cite{FiveGluon}.  On the other hand,
this string theory was not optimized for QCD calculations.
It did not possess all the symmetries of QCD.  For example,
QCD is classically conformally invariant (independent of energy
scale).  But traditional string theory depends on the scale $M_P$,
as reflected in the mass spectrum in \fig{StringSpectrum}(a).
Related to this fact, a fair amount of analysis was required to 
decouple the unwanted massive states from the loop amplitudes 
(as well as the massless states not corresponding to QCD). 
The analysis would have had to be redone to incorporate
external quarks, for example.  In the meantime it was found
that ``abstracting the lessons'' from string theory was often
the most efficient way to proceed.  The efficiency of the
string-based rules for amplitudes could be attributed
to a background-field quantization of gauge theory~\cite{BFGauge}
and a second-order formulation for fermions, for instance~\cite{BDBFGauge}.  
These lessons could then be applied to amplitudes
with external quarks, without having to develop the full
string-theoretic machinery~\cite{QQGGG}.

In contrast, the twistor string theory invented by
Witten~\cite{WittenTopologicalString} is a topological one,
and the string moves in twistor-space, not the usual space-time.
``Topological'' means that the energy of the string
only depends on topological information, so that very few of 
its degrees of freedom are dynamical.  As a result, it does 
not have a tower of massive states, only massless ones, as shown in
\fig{StringSpectrum}(b).   Twistor string theory is conformally
invariant, like classical QCD.  It makes much more manifest
the symmetries of classical QCD, which include not only 
conformal invariance, and the secret ($\Neqfour$) supersymmetry
mentioned in \sect{TwistorSection}, but a full superconformal
group containing them.  So, from the point of view of calculating
QCD amplitudes, twistor string theory seems almost designed to do 
the job.

On the other hand, twistor string theory is still not precisely
QCD.  It possesses all the $\Neqfour$ superpartners of the gluons,
instead of quarks.  It also contains gravitons, but not those 
of Einstein's theory of gravity; instead they belong to
a non-unitary theory, conformal supergravity.  Both of these
properties are not really an issue for computations of tree-level
amplitudes, but they can play havoc with a loop-level description.
In fact, there is no satisfactory one-loop formulation of 
twistor string theory at present.
Once again, however, from a computational point of view,
abstracting the lessons is often the best.

Even at tree level, such abstraction can be beneficial.
Although the MHV $n$-gluon amplitudes~(\ref{PTAmps}) could be 
evaluated directly from the twistor
string~\cite{WittenTopologicalString}, 
and the six-gluon non-MHV amplitudes, such as \eqn{ApppmmmOLD}, 
were also produced in this way~\cite{RSVI},
the MHV rules~\cite{CSWI} have provided a much more efficient method
for generic tree amplitudes.   They originated at least in part
from abstracting the twistor structure which was found by studying existing
QCD amplitudes.  (One could also say, however, that the MHV rules 
follow from a different, ``disconnected'', prescription for evaluating
the relevant twistor-string contributions.)


\section{On-shell recursive methods}
\label{OSRRSection}

Another process of abstraction and streamlining led, at the beginning
of this year, to the on-shell recursion relations of Britto, Cachazo, Feng
and Witten~\cite{BCFRecursive,BCFWRecursive}.  These relations are
even more efficient, and lead to more compact formulas, than the MHV
rules.  Also, they can be proven in a very simple way, using only 
Cauchy's theorem and factorization properties.  So it is very easy
to extend these relations to more general processes, and also to apply
the same kinds of techniques to the computation of one-loop amplitudes
in QCD.

The path to the on-shell recursion relations was somewhat roundabout,
proceeding through the one-loop amplitudes in $\Neqfour$ super-Yang-Mills
theory, whose box coefficients were sketched in \sect{TwistorLoopSection}.
These amplitudes have infrared divergences, represented in dimensional 
regularization as poles in $\e = (4-D)/2$.  The residues
of the poles have to be proportional to the corresponding tree amplitude.
This requirement gave new formulas for tree amplitudes, in terms
of sums of box coefficients~\cite{NMHVSeven,Neq4NMHV,RSVDissolve},
which were more compact than previously-known expressions.
Using generalized unitarity, these formulas could be reinterpreted
as quadratic recursion relations~\cite{BCFRecursive}

\begin{figure*}[t]
\centering
\includegraphics[height=.22\textheight]{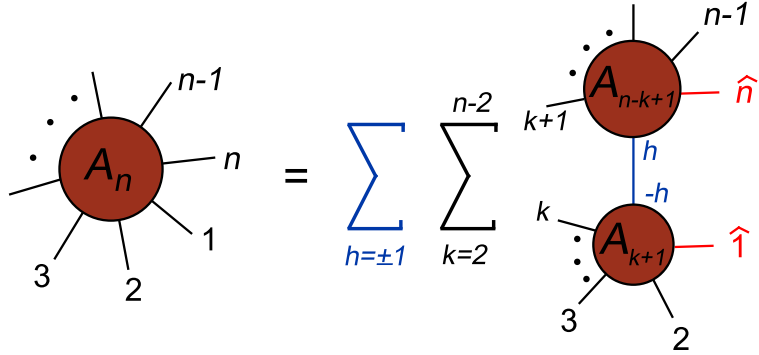}
\caption{Diagrammatic representation of an on-shell recursion 
relation for tree amplitudes.} 
\label{onshellrectreefigure}
\end{figure*}

The basic on-shell recursion relation for tree amplitudes 
reads~\cite{BCFRecursive,BCFWRecursive},
\begin{equation}
A_n^\tree(1,2,\ldots,n) =
\sum_{h=\pm1} \sum_{k=2}^{n-2}
A_{k+1}^\tree(\hat{1},2,\ldots,k,-\hat{K}_{1,k}^{-h})
{i\over K_{1,k}^2}
A_{n-k+1}^\tree(\hat{K}_{1,k}^h,k+1,\ldots,n-1,\hat{n}).
\label{OSRR}
\end{equation}
It is depicted diagrammatically in \fig{onshellrectreefigure}.
The amplitude is represented as a sum of products of lower-point 
amplitudes, evaluated on shell, but for complex, shifted values 
of the momenta (see below).   The helicity labels of the $n$ 
external gluons have been omitted, but they are the same on the 
left- and right-hand sides of \eqn{OSRR}.  
For the relation to be valid, the helicities of gluons $n$ and $1$
can be $(h_n,h_1) = (-1,1)$, $(1,1)$, or $(-1,-1)$, but not $(1,-1)$.
There are two sums.
The first is over the helicity $h$ of an intermediate gluon 
propagating (downward) between the two amplitudes.
The second sum is over an
integer $k$, which labels the different ways the set $\{1,2,\ldots,n\}$
can be partitioned into two cyclicly-consecutive sets, each containing at
least 3 elements, where labels $1$ and $n$ belong to different sets.
A hat on top of a momentum label denotes that the corresponding momentum
is {\it not} that of the original $n$-point amplitude, but is 
shifted to a different value. 

To describe the shifted momenta, first note that, from \eqn{kfact}, 
$k_i^\mu 
= \sigma^\mu_{\alpha\dot\alpha} 
\lambda_i^\alpha \tlambda_i^{\dot\alpha}$ 
is a massless four-vector because of the antisymmetry
of the spinor products,
\begin{equation}
k_i^2 = \ve_{\beta\alpha} \ve_{\dot\alpha\dot\beta} 
  (\ksl_i)^{\alpha\dot\alpha} (\ksl_i)^{\dot\beta\beta}
 =  \ve_{\beta\alpha} \lambda_i^\alpha \lambda_i^\beta 
    \ve_{\dot\alpha\dot\beta}  
     \tlambda_i^{\dot\alpha} \tlambda_i^{\dot\beta}
 = - \spa{i}.{i} \spb{i}.{i} = 0 \,. 
\label{ksqvanish}
\end{equation}
It will continue to be massless even if one of the two spinors is
shifted so that it is no longer the complex conjugate of the
other spinor, for example
\begin{equation}
\hat{k}_i^\mu (\sigma_\mu)_{\alpha\dot\alpha}
= (\hat{\ksl}_i)_{\alpha\dot\alpha}
= (\hat{\lambda}_i)_\alpha (\tlambda_i)_{\dot\alpha} \,,
\label{kfactcx}
\end{equation}
where $\hat{\lambda}_i$ is shifted away from $\lambda_i$.

The momentum shift in the $k^{\rm th}$ term in~\eqn{OSRR} can now be 
described as, 
\begin{eqnarray}
 \lambda_1 &\to& \hat{\lambda}_1 \equiv \lambda_1 + z_k \lambda_n, \qquad\quad
  \tlambda_1 \to \tlambda_1,
 \nonumber \\
 \lambda_n &\to& \lambda_n, \hskip 3cm
  \tlambda_n \to \hat{\tlambda}_n \equiv \tlambda_n - z_k \tlambda_1 \,,
\label{spinorshift1n}
\end{eqnarray}
where
\begin{equation}
z_k = - { K_{1,k}^2  \over \langle n^- | \Ksl_{1,k} | 1^-\rangle } \,.
\label{zk}
\end{equation}
This shift keeps 
$\hat{\ksl}_1 = (\lambda_1 + z_k \lambda_n) \tlambda_1$
and 
$\hat{\ksl}_n = \lambda_n (\tlambda_n - z_k \tlambda_1)$
massless, as discussed above.
It preserves overall momentum conservation, because
$\hat{\ksl}_1 + \hat{\ksl}_n 
= \lambda_1 \tlambda_1 + \lambda_n \tlambda_n 
= \ksl_1 + \ksl_n$.
And the intermediate gluon momentum, defined by 
$\hat{\Ksl}_{1,k} = \Ksl_{1,k} + z_k \lambda_n \tlambda_1$,
is also massless (on shell), because
\begin{equation}
\hat{K}_{1,k}^2 = (\Ksl_{1,k} + z_k \lambda_n \tlambda_1)^2
 = K_{1,k}^2 + z_k \langle n^- | \Ksl_{1,k} | 1^-\rangle
 = 0 \,.
\label{K1knull}
\end{equation}

The derivation of \eqn{OSRR} is very simple~\cite{BCFWRecursive}.
The momentum shift~(\ref{spinorshift1n}) is considered for an arbitrary
complex number $z$, instead of the discrete values $z_k$ in \eqn{zk}.
This shift defines an analytic function of $z$, $A_n^\tree(z)$. 
It has poles in $z$ whenever a collection of the shifted momenta,
corresponding to an intermediate state, can go on shell.  For every
allowed partition of 
$\{1,2,\ldots,n\}$ into $\{1,2,\ldots,k\} \cup \{k+1,\ldots,n-1,n\}$,
there is a unique value of $z$ that accomplishes this, $z_k$, because
$\hat{K}_{1,k}^2(z_k) = 0$ according to \eqn{K1knull}.  
The desired amplitude is the value of $A_n^\tree(z)$ at $z=0$. 
Provided that $A_n^\tree(z) \to 0$ as $z\to\infty$, 
this value at $z=0$ is determined by Cauchy's theorem in terms of 
the residues of $A_n^\tree(z)$ at $z=z_k$.
Using general factorization properties of tree amplitudes,
the $k^{\rm th}$ residue evaluates to the product found in
the $k^{\rm th}$ term in \eqn{OSRR}.
The vanishing of $A_n^\tree(z)$ as $z\to\infty$ can be established
directly from Feynman diagrams for $(h_n,h_1) = (-1,1)$~\cite{BCFWRecursive}.
For the other two valid cases it can be shown using the 
``MHV rules''~\cite{CSWI},
or by a recursive argument~\cite{BGKS}.

No knowledge of twistor space is needed to implement \eqn{OSRR}.
Its derivation is heuristically related to twistor space, however, in 
that spinors, not vectors, play the fundamental role. 

Off-shell recursive approaches to summing Feynman diagrams have a long
history~\cite{BGRecursive,Mahlon}.  In the off-shell case, however,
the auxiliary lower-point quantities are gauge-dependent.
In on-shell recursion relations, in contrast, they are precisely 
the desired physical, gauge-invariant, 
on-shell scattering amplitudes, just with fewer partons.  
In short, trees are recycled into trees.

\subsection{A simple application: $A_6^\tree(1^+,2^+,3^+,4^-,5^-,6^-)$}
\label{ApplicationSubsection}

\begin{figure*}[t]
\centering
\includegraphics[height=.22\textheight]{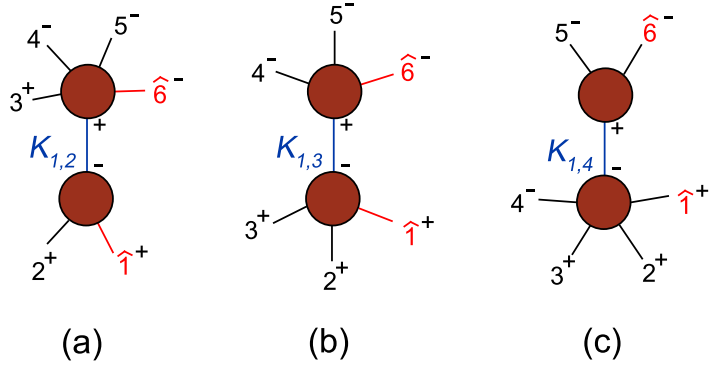}
\caption{On-shell recursive diagrams for 
$A_6^\tree(1^+,2^+,3^+,4^-,5^-,6^-)$.}
\label{BCFpppmmmfigure}
\end{figure*}

Let us now work through a
simple application of~\eqn{OSRR}~\cite{BCFRecursive}.
The first non-MHV $n$-gluon amplitudes (taking into account parity)
are those with six gluons, three of positive helicity and three negative.
There are three cyclicly-inequivalent helicity configurations: 
$A_6^\tree(1^+,2^+,3^+,4^-,5^-,6^-)$, 
$A_6^\tree(1^+,2^+,3^-,4^+,5^-,6^-)$,
and $A_6^\tree(1^+,2^-,3^+,4^-,5^+,6^-)$.  
The last of these amplitudes is related to the first two a 
``dual Ward identity'' (group theory relation)~\cite{TreeReview}.
Here we apply \eqn{OSRR} to $A_6^\tree(1^+,2^+,3^+,4^-,5^-,6^-)$.
Instead of 220 Feynman diagrams (including all color-orderings), 
there are just three potential on-shell recursive diagrams,
shown in \fig{BCFpppmmmfigure}.  Diagrams
of the form of \fig{BCFpppmmmfigure}(a) and \fig{BCFpppmmmfigure}(c),
but with a reversed helicity assignment to the intermediate gluon,
vanish because $A_3^\tree(+,+,+) = A_3^\tree(-,-,-) = 0$.
\Fig{BCFpppmmmfigure}(b), and the
corresponding diagram with a reversed intermediate helicity,
both vanish using \eqn{susyvanish}.  Finally, diagram (c) is
related to diagram (a) by the ``flip''
symmetry $(1\lr6,2\lr5,3\lr4)$ (plus spinor conjugation).

So, remarkably, there is only one independent diagram, 
\fig{BCFpppmmmfigure}(a). 
Its value is given by the product of two shifted MHV amplitudes, 
each parity-conjugated with respect to \eqn{PTAmps},
\begin{eqnarray}
D^{(a)} &=& A_3^\tree(\hat{1}^+,2^+,-\hat{K}_{1,2}^-) 
 { i\over K_{1,2}^2 }
  A_5^\tree(\hat{K}_{1,2}^+,3^+,4^-,5^-,\hat{6}^-)
\nonumber\\
 &=& {-i \over s_{12}}
 { {\spbsh{\hat{1}}.{2}}^3 
   \over \spbsh{2}.{\hat{K}_{1,2}} \spbsh{\hat{K}_{1,2}}.{\hat{1}} }
 { {\spbsh{\hat{K}_{1,2}}.{3}}^3 
   \over \spb3.4 \spb4.5 \spbsh{5}.{\hat{6}} 
                       \spbsh{\hat{6}}.{\hat{K}_{1,2}} } 
\nonumber\\
&=& 
i \,  
{ {\spab{6}.{(1+2)}.{3}}^3
 \over \spa6.1 \spa1.2 \spb3.4 \spb4.5 
      s_{612} \spab{2}.{(6+1)}.{5} } \,.
\label{Davalue}
\end{eqnarray}
To get the contribution $D^{(c)}$, we add the image of $D^{(a)}$
under the permutation $(1\lr6,2\lr5,3\lr4)$, 
combined with spinor conjugation, $\langle\ \rangle \, \lr [\ ]$.
The full amplitude is
\begin{eqnarray}
A_6^\tree(1^+,2^+,3^+,4^-,5^-,6^-)
&=&  i \biggl[ { { \langle 6^- |(1+2) | 3^- \rangle }^3
   \over \spa6.1 \spa1.2 \spb3.4 \spb4.5 s_{612}
\langle 2^- |(6+1)|5^-\rangle }
\nonumber\\
&&\hskip0cm
 + { { \langle 4^- |(5+6) | 1^- \rangle }^3
   \over \spa2.3 \spa3.4 \spb5.6 \spb6.1 s_{561}
\langle 2^- |(6+1)|5^-\rangle } \biggr] \,.~~~~
\label{pppmmmsimple}
\end{eqnarray}
Let's compare this representation of
the amplitude with the previous one, \eqn{ApppmmmOLD},
found using Feynman diagrams.  The second expression
is shorter. (Here the difference in length is minimal; it 
becomes more striking for seven gluons~\cite{BGKSeven,NMHVSeven}.)
It also makes manifest the square-root collinear behavior in 
all channels.  For example, in the collinear limit where $k_3$ becomes
parallel to $k_4$, \eqn{pppmmmsimple} has the correct $1/\spa3.4$
and $1/\spb3.4$ behavior manifest; in \eqn{ApppmmmOLD}, cancellations
between the three terms, each of which behaves like $1/s_{34}$,
are required to obtain the proper behavior.
On the other hand, \eqn{pppmmmsimple} contains a spurious singularity,
because $\langle 2^- |(6+1)|5^-\rangle$ vanishes when 
$k_6+k_1$ happens to be a linear combination of $k_2$ and $k_5$
(use the massless Dirac equation to see this).
The amplitude is perfectly finite in this region, but each term diverges.
In numerically implementing \eqn{pppmmmsimple}, one should take care
in this region.


\section{One-loop amplitudes}
\label{LoopSection}

Although the MHV and on-shell recursive rules are quite efficient 
for the analytical computation of many types of tree amplitudes, 
and shed a lot of light on their structure,
in the end all one really wants are numerical values.  
Quite efficient numerical computer programs 
have already been developed over the years, based on off-shell
recursive methods~\cite{RecursivePrograms}, which can evaluate
QCD tree amplitudes with of order 10 external partons in a reasonable
amount of time.  In contrast, the complete set of one-loop
helicity amplitudes is not known for any pure QCD process with
greater than five external legs.  There are similar bottlenecks
for processes in which a few electroweak vector bosons
are produced in addition to multiple QCD partons.  So it is
of great interest to see whether new methods can be developed
for one-loop QCD amplitudes.

The method used to prove the tree-level on-shell recursion 
relations~\cite{BCFWRecursive} --- shifting a pair of momenta
by a complex amount, while keeping them on shell --- 
is particularly promising in this regard, because it efficiently
incorporates the known factorization of amplitudes onto
collinear and multi-particle poles.  Indeed, the same techniques
can be adopted at one loop, 
in order to determine the rational
(non-logarithmic) parts of amplitudes, once the parts containing
branch cuts (logarithms, polylogarithms, {\it etc.}) have been
determined by other means --- for example, using unitarity,
as mentioned in \sect{TwistorLoopSection}.
Recently, all the one-loop $n$-gluon helicity amplitudes in QCD with up 
to two (adjacent) negative-helicity gluons, and an arbitrary number
of positive helicity ones, have been produced (or reproduced) 
in this way~\cite{BDKLoopFinite,Bootstrapping}. 
The amplitudes having $n_-=0$ or 1 are quite special, because
they vanish at tree-level (\eqn{susyvanish}).
They have no infrared or ultraviolet divergences, and there are
no branch cuts at all.  Also, they were known from previous
work~\cite{BCDK,Mahlon}.  They have a purely recursive representation,
whose construction involved a few assumptions, which could be
cross-checked by comparing to the previous results~\cite{BDKLoopFinite}.

The series of amplitudes $A_n^{\rm 1-loop}(1^-,2^-,3^+,4^+,\ldots,n^+)$,
with two adjacent negative helicities, have branch cuts
as well as infrared and ultraviolet divergences.  The branch cuts
were determined a decade ago, using unitarity~\cite{Fusing}.
The rational parts can now be constructed 
recursively~\cite{Bootstrapping}.
In addition to a set of recursive diagrams, much like the tree-level
formula~(\ref{OSRR}), there are certain ``overlap'' diagrams,
which perform bookkeeping with respect to certain rational-function
terms which naturally accompany the logarithmic terms.
There are relatively few diagrams to evaluate.  For 
the rational part of $A_6^{\rm 1-loop}(1^-,2^-,3^+,4^+,5^+,6^+)$,
for example, there are four nonvanishing recursive diagrams
and three nonvanishing overlap diagrams. The evaluation of each
diagram is completely algebraic; no loop integrations are required.
In contrast, the number of one-loop 6-gluon Feynman diagrams
in QCD is 10,680, each of which requires a loop integration. 

\section{Conclusions}
\label{ConclusionsSection}

Several new methods for computing gauge theory scattering amplitudes
relevant for LHC physics have been developed over the last year 
or two, with a strong stimulation from twistor string theory.  
After some abstraction and streamlining, however, many of these
methods actually bear a close resemblance to the 
bootstrap program developed in the 1960s.  In a bootstrap, scattering
amplitudes are reconstructed directly from their analytic properties, 
without the need for a Lagrangian~\cite{Bootstrap,ELOP}.  
While this program has proven difficult, if not impossible, 
to carry out in full nonperturbative generality in a strongly-coupled
four-dimensional field theory, in the context of perturbation theory
much more information is available to assist it.
The (factorization) poles of amplitudes are dictated
by amplitudes with fewer legs, while the (unitarity) cuts
are dictated by products of amplitudes with fewer loops.
Tree-level on-shell recursion relations, 
for example, are a very convenient way of systematically 
incorporating the factorization data.
The use of analyticity fell somewhat out of favor in the the 
1970s, with the rise of a Lagrangian (QCD) for the strong 
interactions.  Ironically, it now proves useful to resurrect 
analyticity, and a perturbative bootstrap, as a tool for 
computing complicated QCD amplitudes --- for which a direct 
Lagrangian approach, that is, using Feynman rules, can 
be very cumbersome.

To date, the ``practical'' spinoffs from twistor-inspired methods
have been primarily for tree amplitudes (which can also be obtained
by other, numerical methods),
and for loop amplitudes in supersymmetric theories.
But recently, new one-loop helicity amplitudes in full QCD 
have begun to fall to these methods,
suggesting that soon there will be direct phenomenological
applications.  In addition, the recent rapid progress in developing
new computational approaches along these lines is very likely to continue.


\begin{theacknowledgments}
I thank the organizers of the HEP2005 Europhysics Conference
for inviting me to present this talk, and for putting together
a very stimulating meeting.
\end{theacknowledgments}


\end{document}